%% file: main.tex
\documentclass[a4paper,11pt]{article}
\usepackage{jinstpub} 
\usepackage{lineno}
\usepackage{array}
\usepackage{siunitx}

\title{Estimation of waveform deformation with the matched filter}

\author{M. Cappelli,\note{Corresponding author.}}
\author{G. Del Castello}
\author{and M. Vignati}
\affiliation{Dipartimento di Fisica, Sapienza Universita di Roma and INFN Sezione di Roma,
\\Piazzale A. Moro 2, Roma I-00185, Italy}

\emailAdd{matteo.cappelli@roma1.infn.it}

\abstract{In many particle physics experiments the data processing is based  on the analysis of the digitized waveforms provided by the detector. While the waveform amplitude is usually correlated to the event energy, the shape may carry useful information for event discrimination. Thanks to the high signal to noise ratio they provide, matched filters are often applied. Their original design is however intended for the estimation of the waveform amplitude only.

In this work we introduce an analytical extension of the original matched filter for the estimation of a possible shape deformation with respect to a reference template. The new filter is validated on simulations and, with respect to shape parameters calculated on unfiltered waveforms or derived from the original matched filter, it improves the discrimination capability by at least a factor 2 both at low and high signal to noise ratios, making it applicable to the data of several experiments. }


\begin{document}
\maketitle
\flushbottom

\section{Introduction}
\label{sec:introduction}

Particle detectors convert the physical interaction inside the target material into an output electrical signal. In order to enhance the signal to noise ratio (SNR) and to extract information efficiently, the digitized waveforms are often processed offline by applying digital filters~\cite{hamming1998digital}. In particular, matched filters~\cite{Gatti:1986cw, RADEKA196786} are often exploited as they provide an accurate and robust estimation of the signal amplitude and, eventually, of the signal arrival time, in the presence of noise with arbitrary power spectrum. As the pulse amplitude may correlate with the energy of the event, this filtering technique represents often the ideal choice for the energy estimation. 

On the other hand, when the goal is to assess the nature of an event, the attention is focused on the pulse shape~\cite{ROUSH1964112} which can carry information on the type of interaction in the detector. In addition, waveforms whose shape is drastically different from the expected one are probably not generated by a particle interaction, but they could be due to electronic noise or detector instabilities. Pulse shape analysis can hence be used to identify the events of interest (signal) and reject the others (background), whether they are of particle or non-particle origin. 

Pulse shape parameters are used to characterize the waveform. They can be evaluated on the unfiltered data, like in the case of rise and decay times~\cite{arnaboldi2011novel, kuchnir1968time}, or can be derived after processing with a digital filter, such as the matched filter~\cite{Domizio_2011}. When dealing with the raw waveform, the limiting factor in the evaluation of pulse shape parameters is the noise, that lowers the precision. When dealing with filtered waveforms, the limiting factor is the filter bandwidth, which may hide the features of different waveforms. These limits affect the discrimination potential, i.e. the capability of isolating the signal from the background, in particular in the case of low signal to noise ratio or small waveform deformation. 

In this work, we expand the original mathematical formalism of the matched filter to estimate not only the amplitude and the time delay of the pulse, but also a possible deformation with respect to a template pulse. The distortion is formalized as a dilatation or contraction of the time by a factor $\varepsilon$, an additional degree of freedom that is used as pulse shape parameter. There is a conceptual difference between this approach and the one followed in the data analysis of gravitational waves experiments, in which the waveform is compared to a bank of different template signals, in order to find the one with the maximum correlation~\cite{babak2013searching, owen1999matched}. In our case we are instead interested in finding any deviation from a given template, without modeling the unwanted signals.

The filter described in this paper is intended to improve the pulse shape discrimination capabilities for background rejection. This is the case for example of experiments looking for rare processes, such as low mass dark matter (DM) interactions~\cite{Essig:2022dfa}, coherent elastic neutrino nucleus scattering (CE$\nu$NS)~\cite{Abdullah:2022zue} and neutrinoless double beta decay (0$\nu\beta\beta$)~\cite{doi:10.1146/annurev-nucl-101918-023407}. 

In DM and CE$\nu$NS experiments an extremely low detector energy threshold is required to identify the low energy nuclear recoil signals. Pulse shape parameters are used to clean the energy spectrum from detector artifacts and noise spikes close to threshold (see for example~\cite{abdelhameed2019first, lang2010discrimination, abele2023observation, augier2023ricochet, augier2023results, bonet2024pulse}) but their effectiveness is limited because of the low SNR, which can amount to just a few units. A non-identified excess background approaching the threshold is limiting several experiments in the field~\cite{Adari_2022}.

In 0$\nu\beta\beta$ decay experiments, instead, the SNR can be very high, up to several thousands.  However, a very good energy resolution and nearly zero background are needed for the signal identification. The LEGEND experiment, for example, identifies background events via their different pulse shapes~\cite{Lehnert_2016, zsigmond2020legend}, while  pile-up events are expected to be the ultimate background of the CUPID experiment~\cite{Armatol_2021, huang2021pulse, Ahmine:2023xhg}.

\section{Mathematical formulation}
\label{sec:mathematical_formulation}

\subsection{Matched filter}
\label{subsec:matched_filter}
 
The matched filter is based on the hypothesis that a waveform $w(t)$ can be decomposed as:
\begin{equation}
    w(t) = A \cdot s(t-t_0) + n(t)~,  
    \label{eq:hypo_OF}
\end{equation}
where $s(t)$ is the known template signal with unitary amplitude, $n(t)$ is an additive noise component with known power spectrum, $A$ is the unknown signal amplitude and $t_0$ accounts for a possible time delay between the actual signal and the template. Since the noise may have a non-zero autocorrelation, a  $\chi^2$ minimization cannot be performed in the time domain and the estimation of $A$ and $t_0$ is instead performed in the frequency domain~\cite{Gatti:1986cw}. This can be accomplished by defining:
\begin{equation}
    \chi^2(A, t_0) = \int_{-\infty}^{\infty} df \, \frac{|w(f) - A\cdot e^{-i \omega t_0} s(f)|^2}{N(f)}~,
\label{eq:chi2_OF}
\end{equation}
in which $w(f)$ and $s(f)$ are the Fourier transform of $w(t)$ and $s(t)$, respectively, and $N(f)$ is the noise power spectral density. The minimization of this quantity, with respect to the two unknown variables $A$ and $t_0$, provides the best estimators for these latter parameters. Minimizing Eq.~\ref{eq:chi2_OF} with respect to $A$ yields the expression for the filtered waveform, which is a function of the delay $t_0$:
\begin{equation}
     \hat{A}(t_0) = \frac{\widetilde{w}(t_0)}{\sigma^2_s}~,
\label{eq:filtered_OF}
\end{equation}
where:
\begin{equation}
    \widetilde{w}(t_0) = \int_{-\infty}^{\infty} df \, e^{i \omega t_0} \frac{s^*(f)}{N(f)} w(f)~, \qquad \sigma^2_s = \int_{-\infty}^{\infty} df \, \frac{|s(f)|^2}{N(f)}~.
    \label{eq:N_and_sigmaN}
\end{equation}
It can be proved~\cite{golwala2000exclusion} that the best estimator for the pulse amplitude,  $\hat{A}$, is the maximum of the filtered waveform $\hat{A}(t_0)$ in Eq.~\ref{eq:filtered_OF}. Similarly, the time of the maximum of $\hat{A}(t_0)$ is the best estimator of the time delay, that will be referred as $\hat{t}_0$. Finally, the expected amplitude resolution  $\sigma_{A}$ amounts to:
\begin{equation}
    \sigma_{A}^{2} =  \frac{1}{\sigma^{2}_s}~.
\end{equation}

\subsection{Extension to deformed pulses}
\label{subsec:matched_filter_extension}

The only two free parameters in the matched filter are the amplitude $A$ and time delay $t_0$. This implies that the template signal $s(t)$ can only be scaled and time-translated to match the waveform $w(t)$. In order to include a possible deformation of $w(t)$ with respect to $s(t)$, we introduce the following modification to Eq.~\ref{eq:hypo_OF}:
\begin{equation}
    w(t) = A \cdot s[g \cdot (t-t_0)] + n(t)~,
    \label{eq:hypo_ASPECT}
\end{equation}
where the additional degree of freedom $g$ accounts for a possible dilatation ($g < 1$) or contraction ($g > 1$) of the template pulse shape $s(t)$. If we assume that the deformation is small, the equations for the $\chi^2$ minimization simplify considerably, as shown in the following. For small deformation we can write $g$  as:
\begin{equation}
    g = 1 - \varepsilon~,  \quad \varepsilon \approx 0~,
\end{equation}
and then perform a Taylor expansion of the deformed template pulse $s(g \cdot t)$ around $\varepsilon = 0$ as:
\begin{equation}
    s(g \cdot t) \approx s(t) - \varepsilon \cdot \Delta s(t)~, \quad  \Delta s(t) = (t - t_s) \frac{d  s(t)}{d  t}~,
    \label{eq:taylor_expansion}
\end{equation}
in which the subsequent terms are $O(\varepsilon^2)$ and $t_s$ is the ``starting point'' of the deformation. This approximated version of the deformed template pulse is the one that will be correlated to $w(t)$. It has the property that, in the point $t = t_s$, it assumes the same value as the non-deformed template pulse $s(t)$, while at increasing time distances from $t_s$ the deformation increases. In practical applications, $t_s$ can be chosen as the start time of the pulse to ensure a good Taylor approximation.

Starting from Eq.~\ref{eq:hypo_ASPECT} and using the approximation in Eq.~\ref{eq:taylor_expansion}, the frequency domain $\chi^2$ now yields:
\begin{equation}
    \chi^2(A, \varepsilon, t_0) = \int_{-\infty}^{\infty} df \, \frac{|w(f) - A \cdot e^{-i \omega t_0} s(f) + A  \varepsilon \cdot e^{-i \omega t_0} \Delta s(f)|^2}{N(f)}~,
    \label{eq:chi2_ASPECT}
\end{equation}
in which $\Delta s(f)$ is the Fourier transform of the function $\Delta s(t)$.
The procedure is to minimize Eq.~\ref{eq:chi2_ASPECT} with respect to the three unknown variables $A$, $\varepsilon$ and $t_0$. In particular, the estimated value for $\varepsilon$ is intended to be used as shape parameter, measuring the deformation of $w(t)$ with respect to $s(t)$. Minimizing with respect to $A$ and $\varepsilon$ one gets their best estimates, $\hat{A}(t_0)$ and $\hat{\varepsilon}(t_0)$, as a function of the time delay $t_0$:
 \begin{equation}
     \hat{A}(t_0) = \frac{\widetilde{w}(t_0)}{\sigma^2_s} \cdot
     \frac{1 - \rho  \frac{\sigma_s}{\sigma_{\Delta s}} \cdot \frac{\widetilde{\Delta w}(t_0)}{\widetilde{w}(t_0)}}{1-\rho^2} ~,
     \label{eq:A_ASPECT}
 \end{equation}
 \begin{equation}
    \hat{\varepsilon}(t_0) = \frac{
    \rho\, \widetilde{w}(t_0) - \frac{\sigma_s}{\sigma_{\Delta s} } \cdot \widetilde{\Delta w}(t_0) 
    }{
    \frac{\sigma_{\Delta s}}{\sigma_s} \cdot \widetilde{w}(t_0) - \rho\,\widetilde{\Delta w}(t_0)
    }~,
     \label{eq:eps_ASPECT}
 \end{equation}
where:
\begin{align}
    \widetilde{\Delta w}(t_0) = \int_{-\infty}^{\infty} df \, e^{i \omega t_0} \frac{\Delta s^* (f)}{N(f)} w(f)~, \qquad
 \sigma^2_{\Delta s} = \int_{-\infty}^{\infty} df \, \frac{|\Delta s (f)|^2}{N(f)}~, 
\end{align}
and 
\begin{equation}
    \rho  = \frac{1}{\sigma_s \, \sigma_{\Delta s}} \int_{-\infty}^{\infty} df \, \frac{s^*(f)\Delta s(f)}{N(f)}~. 
\end{equation}
The best estimate of the jitter, $\hat{t}_0$, is obtained from the minimum of Eq.~\ref{eq:chi2_ASPECT}, after substituting $A$ and $\varepsilon$ with Eqns.~\ref{eq:A_ASPECT} and \ref{eq:eps_ASPECT}. Then, the best estimators of amplitude and deformation are derived as $\hat{A} = \hat{A}(\hat{t}_0)$ and $\hat{\varepsilon} = \hat{\varepsilon}(\hat{t}_0)$, respectively.

The expected resolution on the  parameters to be fitted, $v_i$, can be derived from the error matrix,
$\sigma^2_{v_i} = (V^{-1})_{ii}$, 
with $V_{ij} = \frac{1}{2}\frac{\partial^2 \chi^2}{\partial v_i \partial v_j}\Bigr|_{\substack{v_i = \hat{v}_i\\v_j = \hat{v}_j}}$.
Therefore, the resolutions on $A$ and $\varepsilon$ are, respectively:
\begin{equation}
    \sigma^2_{A} = \frac{1}{ \sigma^2_s \cdot \left (1 - \rho^2 \right )}~,
\end{equation}
\begin{equation}
    \sigma^2_{\varepsilon} = 
    \frac{\sigma^2_{A}}{\hat{A}^2}\cdot
    \frac{\sigma^2_s - 2\hat{\varepsilon}\sigma_s\sigma_{\Delta s}\rho + \hat{\varepsilon}^2\sigma^2_{\Delta s}}{\sigma^2_{\Delta s}}~.
    \label{eq:sigma_eps}
\end{equation}

\section{Simulation}
\label{subsec:simulation}

We validated our technique on simulated events. We generated pulses with rising and trailing edges described by a single time constant, respectively:
\begin{equation}
    s[t] = k \cdot \theta(t-\bar{t}) \left( -e^{-\frac{t-\bar{t}}{\tau_r}} + e^{-\frac{t-\bar{t}}{\tau_d}} \right )~,
    \label{eq:model}
\end{equation}
where $t$ is the discrete sample index, $\bar{t}$ is the pulse start time in samples, $\tau_r$ and $\tau_d$ are, respectively, the rise and decay constants of the pulse and 
\begin{equation}
    k = \left[ \left(\frac{\tau_r}{\tau_d}\right)^{\frac{\tau_r}{\tau_d-\tau_r}} - \left(\frac{\tau_r}{\tau_d}\right)^{\frac{\tau_d}{\tau_d-\tau_r}} \right]^{-1}
\end{equation}
is a normalization constant needed to have a model with unitary amplitude. The length of the window is chosen as $N = 2000$ samples, and the parameters in Eq.~\ref{eq:model} are chosen as $\bar{t} = 400$, $\tau_r = 10$ and $\tau_d = 100$ samples. A generic waveform hence has the form:
\begin{equation}
    w[t] = A \cdot s[ (1-\varepsilon)\cdot(t-t_0)] + n[t]~, 
    \label{eq:pulse}
\end{equation}
where $A$ is the pulse amplitude, $\varepsilon$ is the deformation, $t_0$ is a time jitter with respect to $\bar t$ and $n[t]$ is a simulated additive noise component with power spectral density following a $f^{-1}$ behaviour and a unitary standard deviation $\sigma$. 

For the data processing, Eqns.~in Sec.~\ref{sec:mathematical_formulation} are ported to the discrete case replacing integrals with sums. In order to build the filter, the template pulse $s[t]$ is estimated back from the generated data by averaging over 500 non-deformed ($\varepsilon = 0$) and non-delayed ($t_0 = 0$) waveforms
and the noise power spectral density, $N[f]$, is estimated from the average over 500 power spectra of generated noise traces. The parameter $t_s$ in Eq.~\ref{eq:taylor_expansion} in principle should be chosen equal to the true start time of the pulse $\bar{t}$. However in real data the start time has to be estimated directly from the template pulse. In this simulation, we set $t_s$ as the point at which the template reaches the 5\% of its maximum amplitude on the rising edge. 

\begin{figure}[tb]
    \centering
    \includegraphics[width = \textwidth]{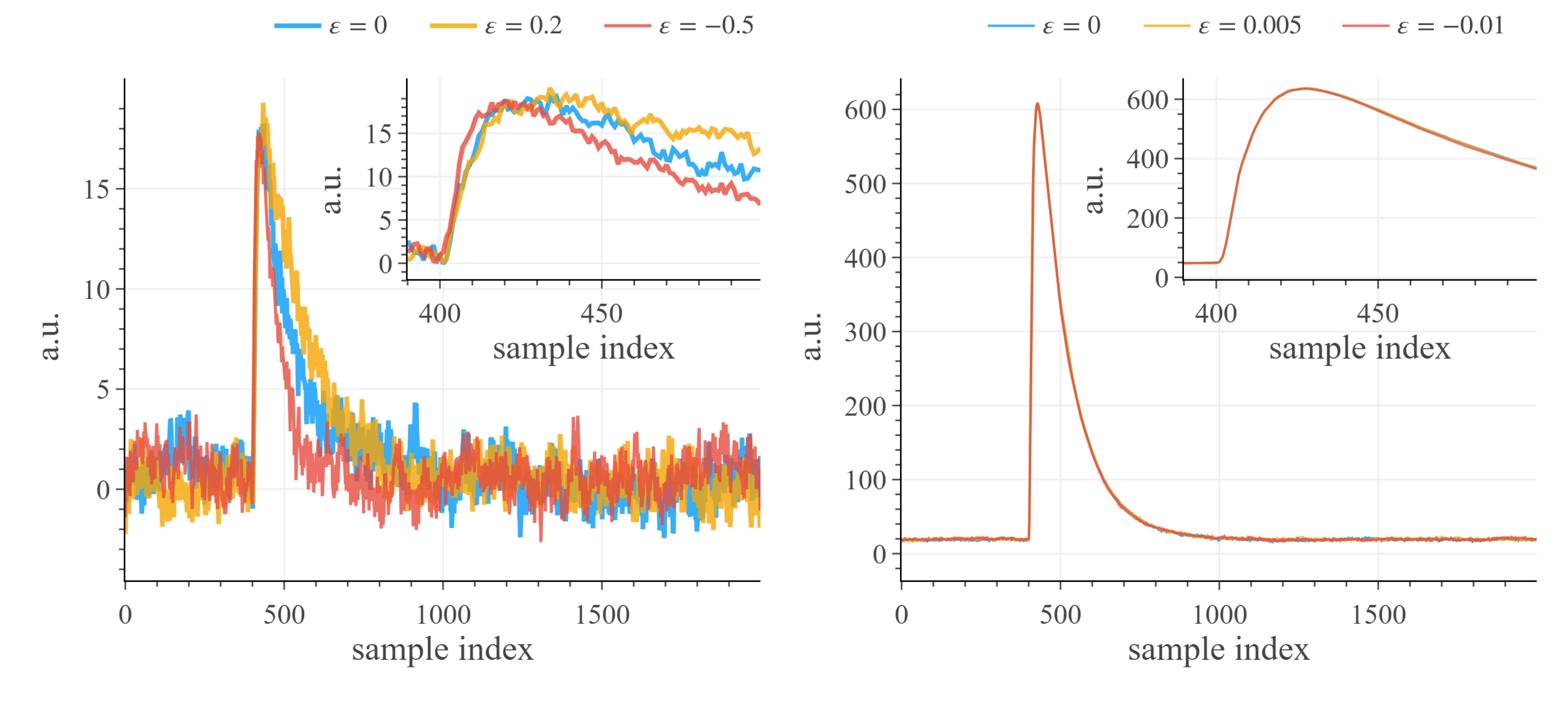}
    \caption{Example of simulated pulses obtained from Eq.~\ref{eq:pulse}. \textbf{Left:} pulses with SNR = 30 and three values for the deformation $\varepsilon$. \textbf{Right:} pulses with SNR = 1000 and three small values for the deformation $\varepsilon$. In this case, no difference between different pulses is visible by eye.}
    \label{fig:pulses}
\end{figure}

In the following we show the results of the application of the filter on different datasets, one featuring low SNR, emulating DM and CE$\nu$NS experiments, and one featuring high SNR, emulating $0\nu\beta\beta$ experiments. The values for the SNR are estimated from the original matched filter as:
\begin{equation}
    \text{SNR} = \frac{A}{\sigma_{A}}~.
    \label{eq:SNR}
\end{equation}
As an example of simulation, Fig.~\ref{fig:pulses} shows simulated waveforms with different values of $\varepsilon$ for $\text{SNR}=30$ (left) and $\text{SNR}=1000$ (right). 

\subsection{Low SNR results}

\begin{figure}[tb]
    \centering
    \includegraphics[width = \textwidth]{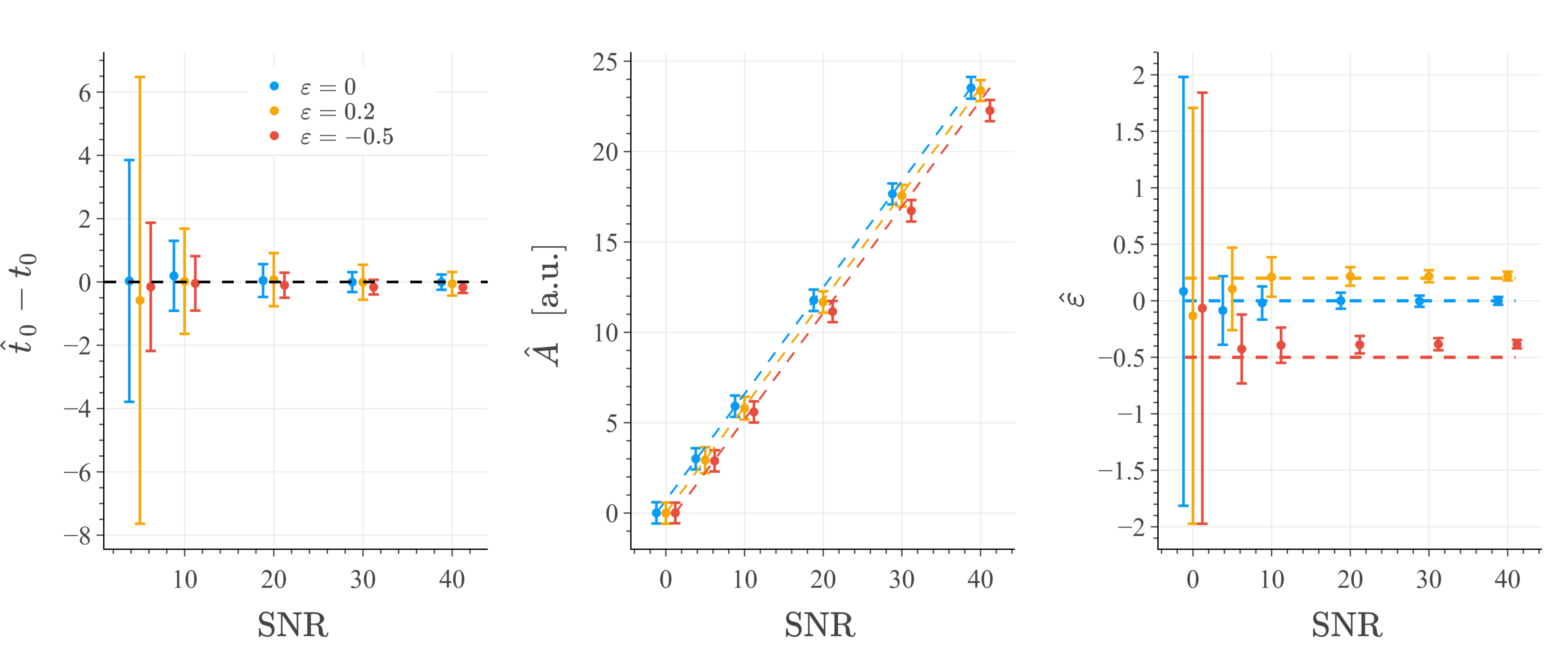}
    \caption{Comparison between the estimates and the true values of the parameters, as a function of the SNR. In each plot, the colors identify different values of the true shape deformation $\varepsilon$. Data points and error bars are, respectively, mean values and standard deviations of the distributions. For display purposes, the points are slightly shifted on x-axis to prevent overlapping. The points are expected to lay on the dashed lines, in the case of a perfect reconstruction. 
    {\bf Left:} difference between the reconstructed and simulated time jitter $\hat{t}_0 - t_0$;
    {\bf Middle:} comparison between the reconstructed  ($\hat{A}$) and simulated ($A$) amplitudes;
    {\bf Right:} reconstructed deformation $\hat{\varepsilon}$.}
    \label{fig:reco_results}
\end{figure}

A first dataset is composed of pulses with $\rm{SNR \leq 40}$ and a deformation like the one introduced in Sec.~\ref{subsec:matched_filter_extension}. Every event in the simulation consists in a waveform of the form in Eq.~\ref{eq:pulse} with  $A$ and $\varepsilon$ chosen among the following values:
\begin{equation}
    \text{SNR} \in \{ 0, 5, 10, 20, 30, 40 \}~, \quad \varepsilon \in \{0, 0.2, -0.5\}~,
\end{equation}
and $t_0$ extracted from an uniform distribution in the range $[-10,10]$ samples.
Figure \ref{fig:reco_results} shows the reconstructed parameters $\hat{t}_0$, $\hat{A}$ and $\hat{\epsilon}$ as a function of the SNR. In the case of events with $\text{SNR} = 0$, which are noise traces, the time jitter is not estimated but we fix $\hat{t}_0 = 0$. In general, the reconstruction is well performed under the hypothesis of low deformation (blue and yellow dots in Fig.~\ref{fig:reco_results} are compatible with the dashed lines representing true values), proving the consistence of the mathematical formulation described in Sec.~\ref{subsec:matched_filter_extension}.

The new parameter $\hat{\varepsilon}$ has been designed to discriminate between the shape of signal  ($\varepsilon = 0$) and background ($\varepsilon \neq 0$) events. In the following we compare its performance to other shape parameters often used for background rejection, such rise time (time interval between 10\% and 90\% of the signal maximum on the rising edge), decay time (time interval between 90\% and 30\% of the signal maximum on the trailing edge) and the $\chi^2$ of the matched filter (Eq.~\ref{eq:chi2_OF}).

\begin{figure}[tb]
    \centering
    \includegraphics[width = \textwidth]{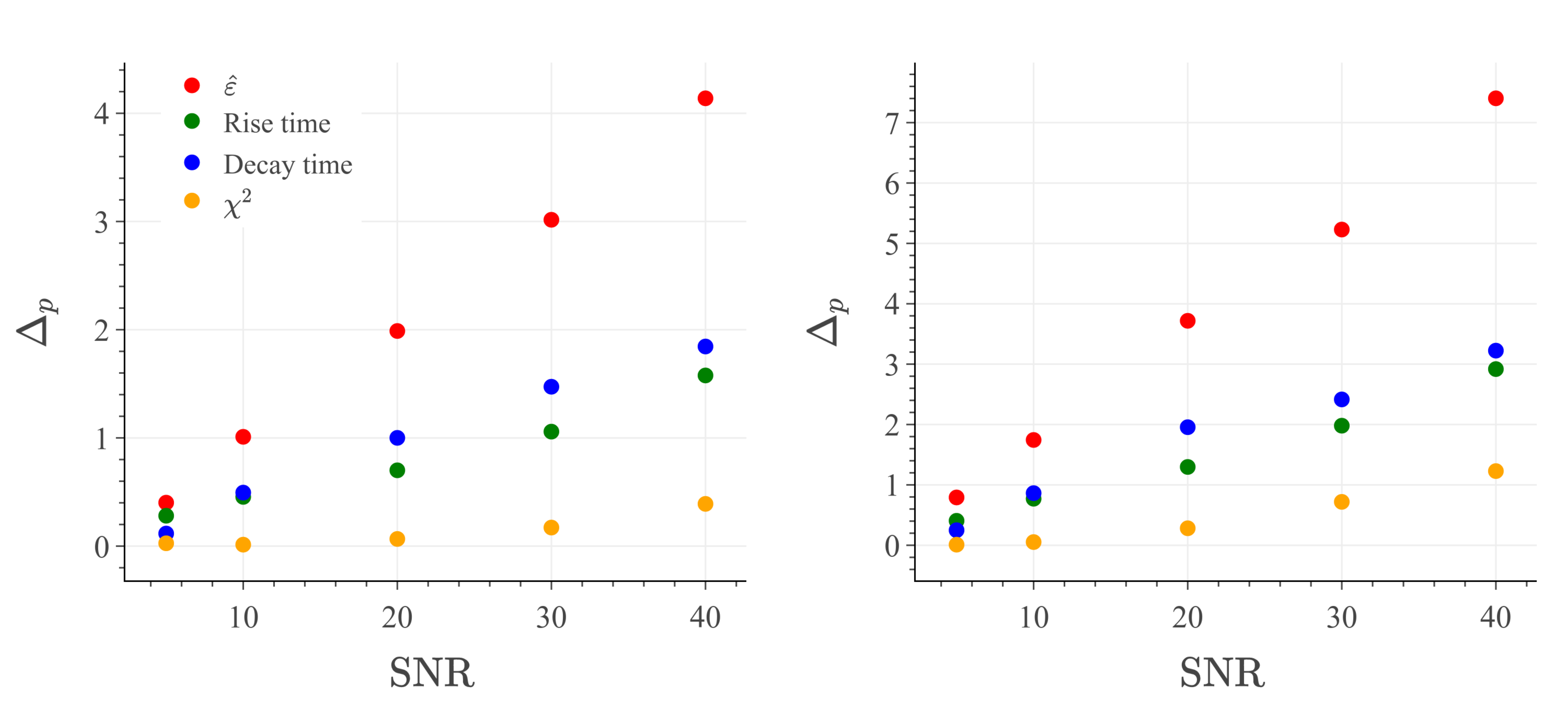}
    \caption{Discrimination potential $\Delta_p(\varepsilon)$ as a function of the SNR, for the  pulse shape parameters $p$ considered in this analysis: rise time, decay time, matched filter $\chi^2$ and reconstructed deformation $\hat{\varepsilon}$ for $\varepsilon = 0.2$ (left) and $\varepsilon = -0.5$ (right).}
    \label{fig:discr_potential}
\end{figure}

In order to compare the performance of different shape parameters $p$ we define the discrimination potential as:
\begin{equation}
    \Delta_p(\varepsilon) = \frac{| <p>_{\varepsilon} - <p>_{\varepsilon = 0} |}{\sqrt{\sigma^2_{p,\varepsilon} + \sigma^2_{p, \varepsilon=0}}}~,
\end{equation}
where $<p>_{\varepsilon}$ and $\sigma^2_{p,\varepsilon}$ are, respectively, the mean value and the variance of the parameter $p$ evaluated on events with a deformation $\varepsilon$. Figure \ref{fig:discr_potential} displays the discrimination potential for the parameters considered, both in the case of low deformation ($\varepsilon = 0.2$) and high deformation ($\varepsilon = -0.5$), showing that our new shape parameter improves the discrimination at low SNR values by a factor 2.5 with respect to the rise time, 2 with respect to the decay time and 10 with respect to the $\chi^2$ of the matched filter.

\begin{figure}[tb]
    \centering
    \includegraphics[width = \textwidth]{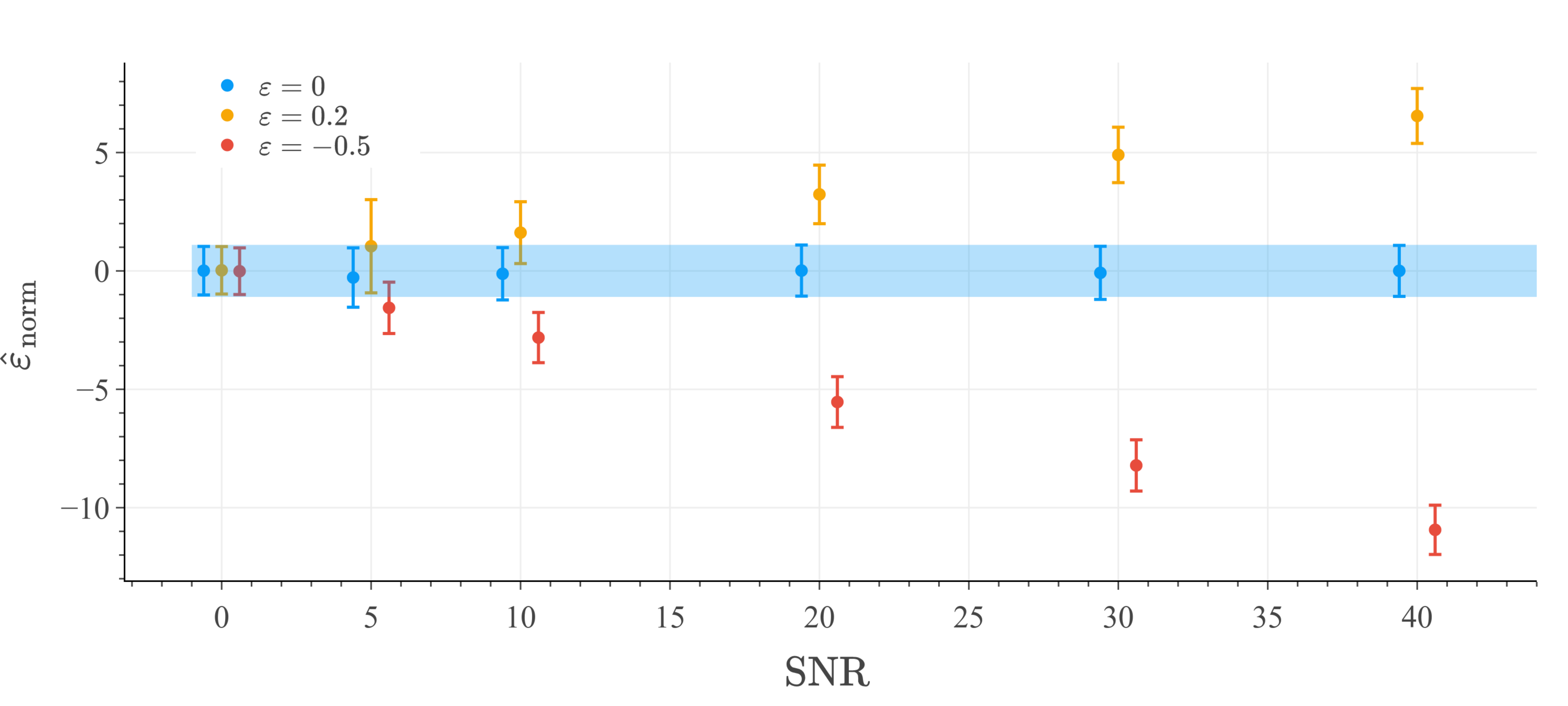}
    \caption{Normalized pulse shape parameter $\hat{\varepsilon}_{\text{norm}}$ as a function of SNR, for different values of $\varepsilon$ (colors). Mean values (points) and standard deviations (error bars) are obtained by performing a Gaussian fit to the distributions. For display purposes, points are slightly shifted on x-axis to prevent overlapping. The blue band represents a possible $\pm 1\sigma$ acceptance region to retain signal  ($\varepsilon=0$) and discard background ($\varepsilon\neq0$).}
    \label{fig:eps_norm}
\end{figure}

The variance of $\hat{\varepsilon}$ depends on the amplitude (see Eq.~\ref{eq:sigma_eps} and Fig.~\ref{fig:reco_results}), and is not of practical use when one needs a selection variable. 
From Eq.~\ref{eq:sigma_eps}, we can define a normalized pulse shape parameter as:
\begin{equation}
    \hat{\varepsilon}_{\text{norm}} = \frac{\hat{\varepsilon}}{\sigma_{\varepsilon}( \varepsilon= 0)}~,
\end{equation}
which has the advantage of having unitary standard deviation, being independent on amplitude and maintaining the same discrimination potential of $\hat{\varepsilon}$. In this way, background events can be rejected with a constant cut that is energy independent, as shown in Fig.~\ref{fig:eps_norm}. 
        
\begin{figure}[tb]
    \centering
    \includegraphics[width = \textwidth]{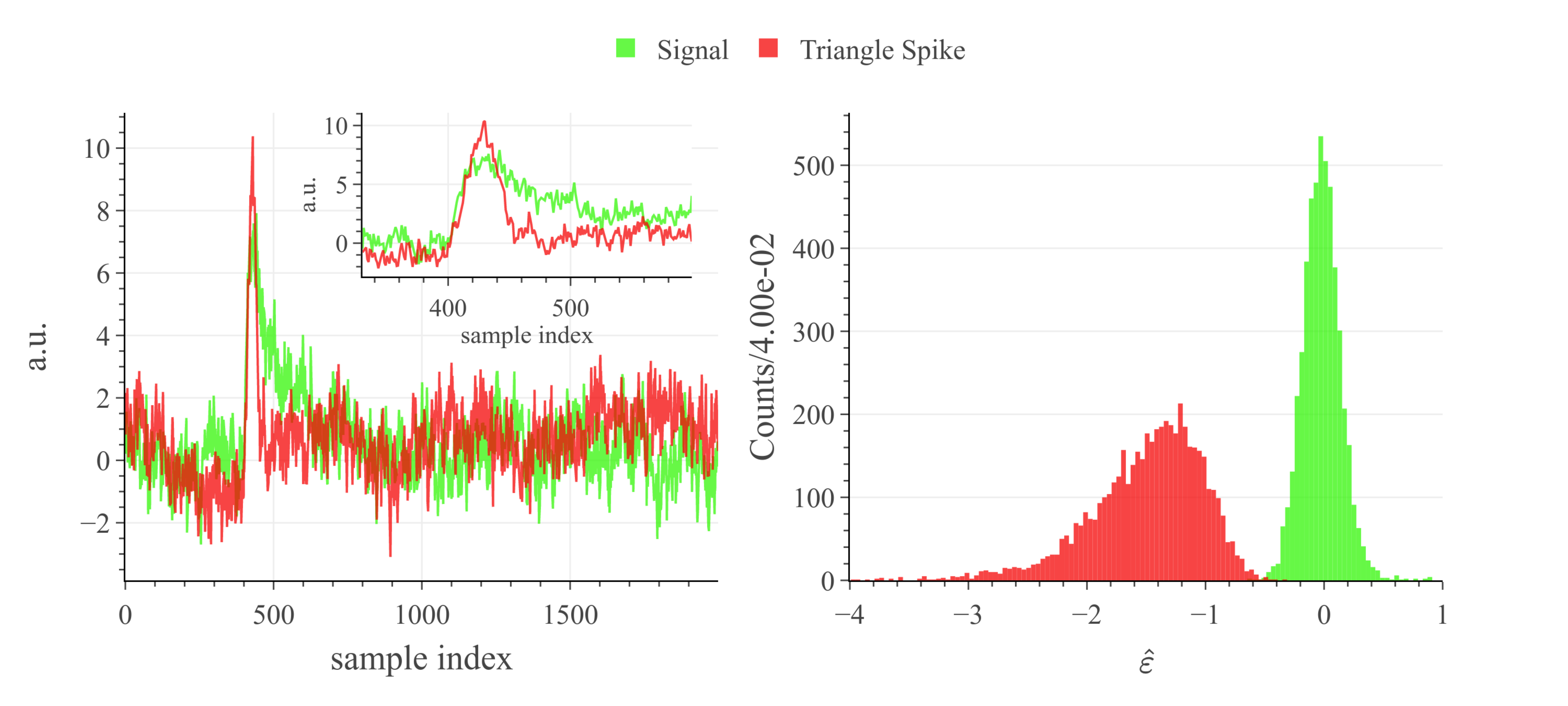}
    \caption{\textbf{Left:} comparison between a signal event and a triangular waveform, both with $\rm{SNR=10}$ after the matched filter. \textbf{Right:} distribution of the parameter $\hat{\varepsilon}$ on events of this kind.}
    \label{fig:noise}
\end{figure}

In order to assess the filter performance on events in which the deformation differs from the expected global time dilation/contraction, we simulated triangular waveforms with $\rm{SNR=10}$ intended to reproduce noise spikes. The length of such spikes is set to 50 samples, which is half of the pulse decay time constant $\tau_d$. A waveform of this kind, together with a signal pulse, is represented in Fig.~\ref{fig:noise}. Also in this case, the pulse shape variable $\hat{\varepsilon}$ provided by the algorithm shows a clear separation between signal and background, as seen by the distributions in Fig.~\ref{fig:noise}. The discrimination potential in this case amounts to $\Delta_{\hat{\varepsilon}}=3.6$, to be compared with $\Delta_{\text{\,decay time}}=3.0$, $\Delta_{\chi^2}=1.4$ and $\Delta_{\text{\,rise time}}=0.7$. 

\subsection{High SNR results}
We studied the performance of the algorithm on events with high SNR and very small deformation:
\begin{equation}
    \text{SNR}=1000~, \quad \varepsilon \in \{0, 0.005, -0.01\}~.
\end{equation}
In this case, $t_0$ is reduced to the range $[-2,2]$ samples in order to emulate a more precise triggering given the higher SNR and for the estimation of $\hat{t}_0$  we interpolate the $\chi^2$ function  around its minimum with a cubic spline in order to reduce biases due to sampling (Eq.~\ref{eq:chi2_ASPECT} ported to the discrete case). 
Then $\hat{A}[t_0]$ and $\hat{\varepsilon}[t_0]$ are also evaluated through a local interpolation of their sampled values. 

\begin{figure}[tb]
    \centering
    \includegraphics[ width = \textwidth]{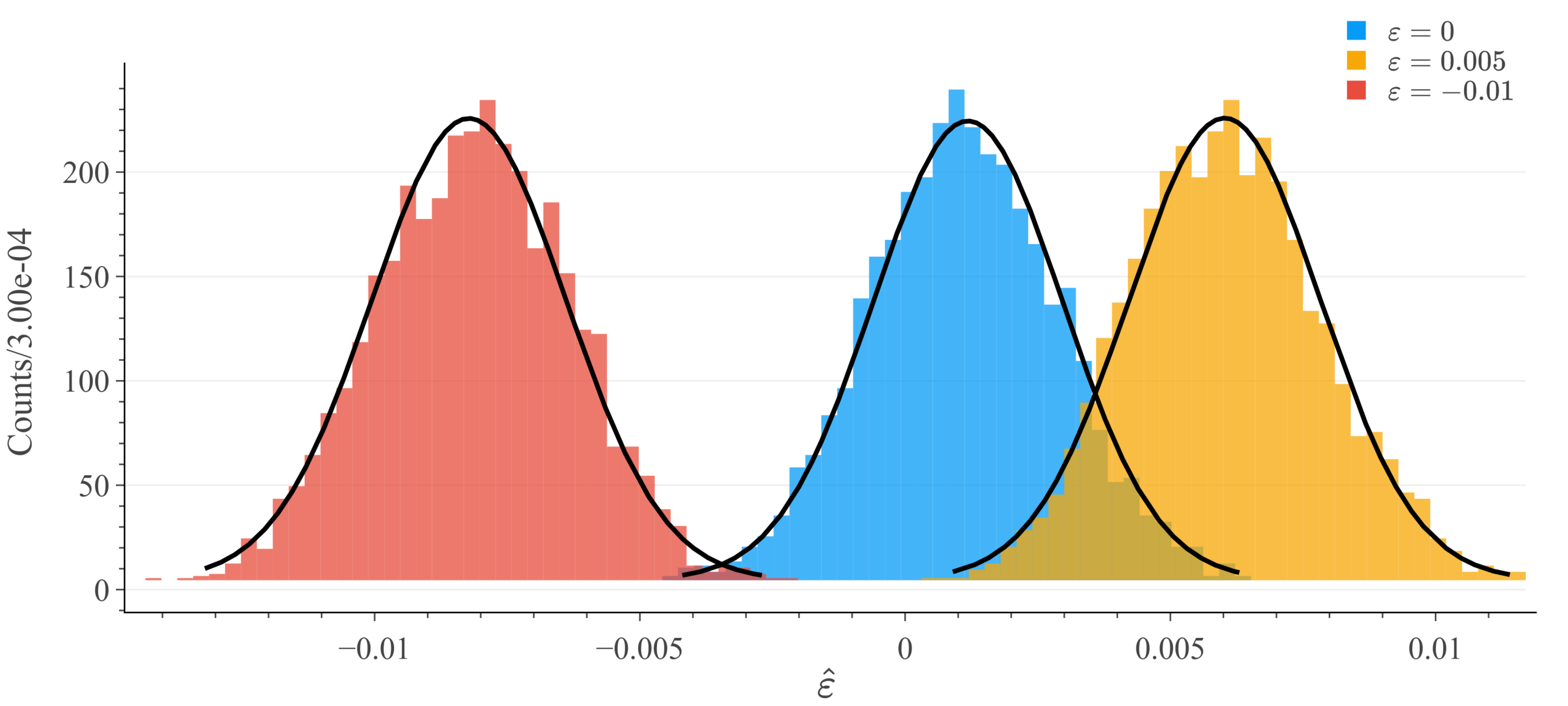}
    \caption{Distribution of the variable $\hat{\varepsilon}$ on events with  SNR~$= 1000$. The blue distribution is for signal events ($\varepsilon = 0$), while the red and the yellow are relative to background events with $\varepsilon \neq 0$. Gaussian fits of the distributions are represented as black solid lines.}
    \label{fig:gaussians}
\end{figure}

Figure \ref{fig:gaussians} shows the distributions of the reconstructed $\hat{\varepsilon}$ in the simulation, for different values of the true deformation $\varepsilon$. A small constant bias in the mean values, possibly still due to sampling, is present and amounts to around $+0.001$. The bias is however roughly the same for the different values of $\varepsilon$ and does not limit the discrimination. A similar bias is observed in the width of the distributions, which result 30 \% higher than their theoretical value from Eq.~\ref{eq:sigma_eps}. A strong separation between signal ($\varepsilon=0$) and background ($\varepsilon\neq0$) is visible, and also in this case the discrimination potential of $\hat{\varepsilon}$ is more than a factor 2 higher than the one of the other parameters (see Tab.~\ref{tab:discr_potential}). 

\begin{table}
    \centering
    \begin{tabular}{|c|l|l|} \hline  
         &   $\Delta_p(\varepsilon = 0.005)$& $\Delta_p(\varepsilon = -0.01)$\\ \hline  
         $\hat{\varepsilon}$& 
     \hspace{0.9 cm}1.9&   \hspace{0.9 cm} 3.8\\ \hline  
 decay time& \hspace{0.9 cm}0.79&  \hspace{0.9 cm} 1.6\\ \hline 
 rise time& \hspace{0.9 cm}0.74&  \hspace{0.9 cm}  1.4\\ \hline  
 $\chi^2$& \hspace{0.9 cm}0.057 & \hspace{0.9 cm} 0.076\\ \hline\end{tabular}
    \caption{Comparison of the discrimination potential $\Delta_p$ of the various parameters $p$ considered in this analysis, for events with SNR = 1000. The first column is for events with deformation $\varepsilon = 0.005$, the second one is for events with deformation $\varepsilon = -0.01$.}
    \label{tab:discr_potential}
\end{table}

\begin{figure}[tb]
    \centering
    \includegraphics[ width = \textwidth]{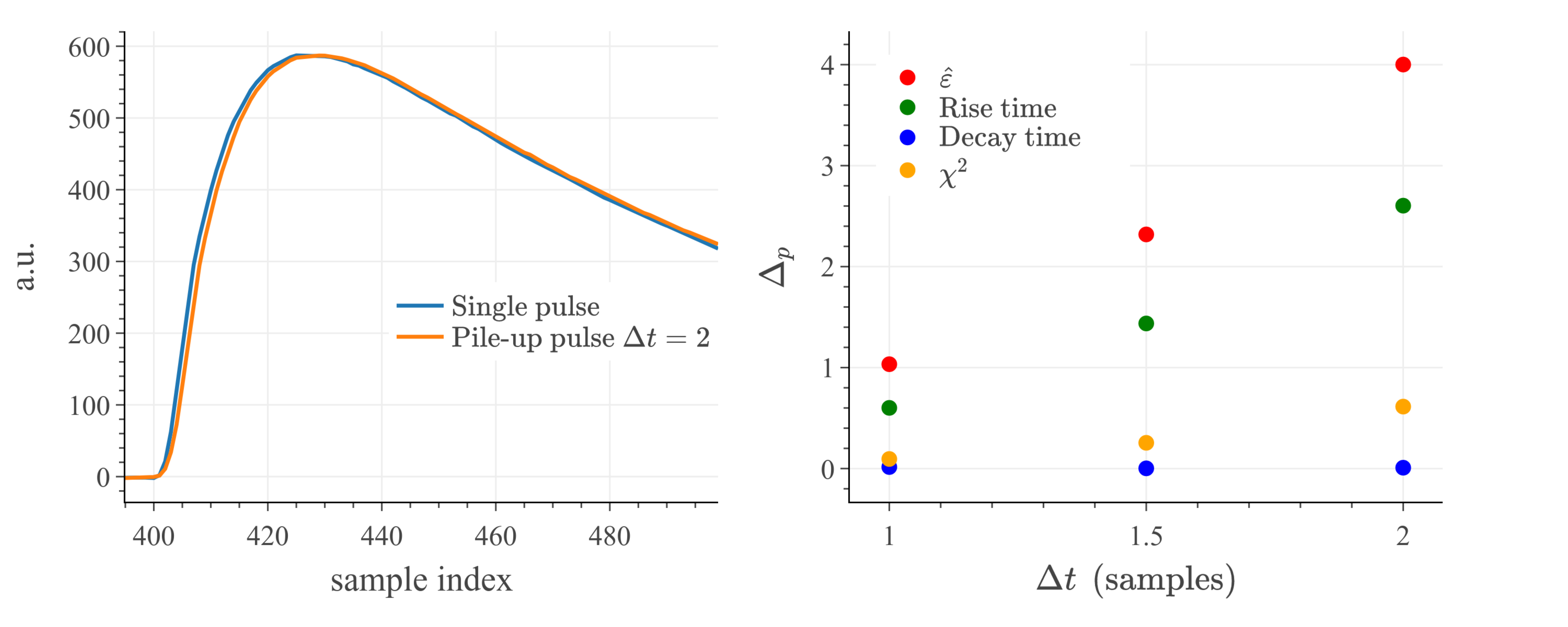}
    \caption{\textbf{Left:} example of a simulated pile-up pulse (zoom on the rising edge), obtained from Eq.~\ref{eq:pile-up} adding two pulses with $\rm{SNR=500}$ and $\Delta t = 2$. For comparison, a single pulse with $\rm{SNR=1000}$ is also shown. \textbf{Right:} discrimination potential $\Delta_p$ as a function of the time separation in pile-up pulses $\Delta t$, for different pulse shape parameters.}
    \label{fig:pileup}
\end{figure}

Finally, we tested the algorithm on pile-up events, simulated by summing up two identical pulses of the form \ref{eq:model} with a time delay $\Delta t$:
\begin{equation}
    w[t] = A \cdot ( s[t] + s[t-\Delta t] ) + n[t]~, 
    \label{eq:pile-up}
\end{equation}
and $A$ chosen in order to have $\rm{SNR=500}$. As before, we are interested on the ability to identify signal events ($\Delta t = 0$) over background events ($\Delta t \neq 0$) with our shape parameter $\hat{\varepsilon}$. We tested small values of the separation,  $  \Delta t \in \{0, 1, 1.5, 2\}$ samples, resulting  in  small differences between single and pile-up pulses, as shown in Fig.~\ref{fig:pileup} (left). The discrimination potential $\Delta_p$ is compared in Fig.~\ref{fig:pileup} (right) to the rise time and the $\chi^2$ parameters. The new shape parameter is once again more powerful in background identification than other variables. 

\section{Conclusions}
\label{sec:conclusions}
In this work we presented an extension of the matched filter intended to reconstruct an eventual deformation of waveforms with respect to a reference template, in addition to the amplitude and the time delay. The minimization used to estimate the best parameters can be easily solved under the hypothesis of low deformation and the expected resolution on the parameters can also be computed. 

We validated the estimation procedure with a Monte Carlo simulation. The simulation proved the validity of the algorithm since the reconstructed quantities are compatible with the true values, when the hypothesis of small deformation is respected. In addition we investigated the possibility to exploit the reconstructed deformation as a new pulse shape parameter, used to identify background events to be rejected. The signal/background separation provided by this new variable exceeds the one of other parameters by at least a factor 2 both at low and high SNR when the deformation is the expected one. This good behaviour is kept also with other kind of background events, like triangular noise spikes and pile-up pulses, making this filtering technique a promising tool for background rejection to be tested in a wide range of experiments.   

\acknowledgments
We thank F.~Pannarale for interesting discussions and the team of the Laboratorio Rivelatori Criogenici of Sapienza U. for its support.
This work was partially supported by Sapienza U. through the grant DANAE-TD and by the European Research Council through the Consolidator Grant ``DANAE'' number 101087663.


\bibliographystyle{JHEP}
\input{main.bbl}

\end{document}

%% file: main.bbl
\providecommand{\href}[2]{#2}\begingroup\raggedright\endgroup

%% file: main.bbl
\begin{thebibliography}{10}

\bibitem{hamming1998digital}
R.W.~Hamming, \emph{Digital filters}, Courier Corporation (1998).

\bibitem{Gatti:1986cw}
E.~Gatti and P.F.~Manfredi, \emph{{Processing the Signals From Solid State Detectors in Elementary Particle Physics}}, \href{https://doi.org/10.1007/BF02822156}{\emph{Riv. Nuovo Cim.} {\bfseries 9N1} (1986) 1}.

\bibitem{RADEKA196786}
V.~Radeka and N.~Karlovac, \emph{Least-square-error amplitude measurement of pulse signals in presence of noise}, \href{https://doi.org/https://doi.org/10.1016/0029-554X(67)90561-7}{\emph{Nuclear Instruments and Methods} {\bfseries 52} (1967) 86}.

\bibitem{ROUSH1964112}
M.~Roush, M.~Wilson and W.~Hornyak, \emph{Pulse shape discrimination}, \href{https://doi.org/https://doi.org/10.1016/0029-554X(64)90333-7}{\emph{Nuclear Instruments and Methods} {\bfseries 31} (1964) 112}.

\bibitem{arnaboldi2011novel}
C.~Arnaboldi, C.~Brofferio, O.~Cremonesi, L.~Gironi, M.~Pavan, G.~Pessina et~al., \emph{A novel technique of particle identification with bolometric detectors}, {\emph{Astroparticle Physics} {\bfseries 34} (2011) 797}.

\bibitem{kuchnir1968time}
F.T.~Kuchnir and F.J.~Lynch, \emph{Time dependence of scintillations and the effect on pulse-shape discrimination}, {\emph{IEEE Transactions on Nuclear Science} {\bfseries 15} (1968) 107}.

\bibitem{Domizio_2011}
S.D.~Domizio, F.~Orio and M.~Vignati, \emph{Lowering the energy threshold of large-mass bolometric detectors}, \href{https://doi.org/10.1088/1748-0221/6/02/p02007}{\emph{Journal of Instrumentation} {\bfseries 6} (2011) P02007–P02007}.

\bibitem{babak2013searching}
S.~Babak, R.~Biswas, P.~Brady, D.A.~Brown, K.~Cannon, C.D.~Capano et~al., \emph{Searching for gravitational waves from binary coalescence}, {\emph{Physical Review D} {\bfseries 87} (2013) 024033}.

\bibitem{owen1999matched}
B.J.~Owen and B.S.~Sathyaprakash, \emph{Matched filtering of gravitational waves from inspiraling compact binaries: Computational cost and template placement}, {\emph{Physical Review D} {\bfseries 60} (1999) 022002}.

\bibitem{Essig:2022dfa}
R.~Essig et~al., \emph{{Snowmass2021 Cosmic Frontier: The landscape of low-threshold dark matter direct detection in the next decade}},  in \emph{{Snowmass 2021}}, 3, 2022 [\href{https://arxiv.org/abs/2203.08297}{{\ttfamily 2203.08297}}].

\bibitem{Abdullah:2022zue}
M.~Abdullah et~al., \emph{{Coherent elastic neutrino-nucleus scattering: Terrestrial and astrophysical applications}},  \href{https://arxiv.org/abs/2203.07361}{{\ttfamily 2203.07361}}.

\bibitem{doi:10.1146/annurev-nucl-101918-023407}
M.J.~Dolinski, A.W.~Poon and W.~Rodejohann, \emph{Neutrinoless double-beta decay: Status and prospects}, \href{https://doi.org/10.1146/annurev-nucl-101918-023407}{\emph{Annual Review of Nuclear and Particle Science} {\bfseries 69} (2019) 219} [\href{https://arxiv.org/abs/https://doi.org/10.1146/annurev-nucl-101918-023407}{{\ttfamily https://doi.org/10.1146/annurev-nucl-101918-023407}}].

\bibitem{abdelhameed2019first}
A.H.~Abdelhameed, G.~Angloher, P.~Bauer, A.~Bento, E.~Bertoldo, C.~Bucci et~al., \emph{First results from the cresst-iii low-mass dark matter program}, {\emph{Physical Review D} {\bfseries 100} (2019) 102002}.

\bibitem{lang2010discrimination}
R.~Lang, G.~Angloher, M.~Bauer, I.~Bavykina, A.~Bento, A.~Brown et~al., \emph{Discrimination of recoil backgrounds in scintillating calorimeters}, {\emph{Astroparticle Physics} {\bfseries 33} (2010) 60}.

\bibitem{abele2023observation}
H.~Abele, G.~Angloher, A.~Bento, L.~Canonica, F.~Cappella, L.~Cardani et~al., \emph{Observation of a nuclear recoil peak at the 100 ev scale induced by neutron capture}, {\emph{Physical Review Letters} {\bfseries 130} (2023) 211802}.

\bibitem{augier2023ricochet}
C.~Augier, G.~Beaulieu, V.~Belov, L.~Berge, J.~Billard, G.~Bres et~al., \emph{Ricochet progress and status}, {\emph{Journal of Low Temperature Physics} (2023) 1}.

\bibitem{augier2023results}
C.~Augier, G.~Baulieu, V.~Belov, L.~Berg{\'e}, J.~Billard, G.~Bres et~al., \emph{Results from a prototype tes detector for the ricochet experiment}, {\emph{arXiv preprint arXiv:2304.14926} (2023) }.

\bibitem{bonet2024pulse}
H.~Bonet, A.~Bonhomme, C.~Buck, K.~F{\"u}lber, J.~Hakenm{\"u}ller, J.~Hempfling et~al., \emph{Pulse shape discrimination for the conus experiment in the kev and sub-kev regime}, {\emph{The European Physical Journal C} {\bfseries 84} (2024) 139}.

\bibitem{Adari_2022}
P.~Adari, A.A.~Aguilar-Arevalo, D.~Amidei, G.~Angloher, E.~Armengaud, C.~Augier et~al., \emph{Excess workshop: Descriptions of rising low-energy spectra}, \href{https://doi.org/10.21468/scipostphysproc.9.001}{\emph{SciPost Physics Proceedings} (2022) }.

\bibitem{Lehnert_2016}
B.~Lehnert, \emph{Background rejection of n+surface events in gerda phase ii}, \href{https://doi.org/10.1088/1742-6596/718/6/062035}{\emph{Journal of Physics: Conference Series} {\bfseries 718} (2016) 062035}.

\bibitem{zsigmond2020legend}
A.J.~Zsigmond, L.~Collaboration et~al., \emph{Legend: The future of neutrinoless double-beta decay search with germanium detectors},  in \emph{Journal of Physics: Conference Series}, vol.~1468, p.~012111, IOP Publishing, 2020.

\bibitem{Armatol_2021}
A.~Armatol, E.~Armengaud, W.~Armstrong, C.~Augier, F.T.~Avignone, O.~Azzolini et~al., \emph{Novel technique for the study of pileup events in cryogenic bolometers}, \href{https://doi.org/10.1103/physrevc.104.015501}{\emph{Physical Review C} {\bfseries 104} (2021) }.

\bibitem{huang2021pulse}
R.~Huang, E.~Armengaud, C.~Augier, A.~Barabash, F.~Bellini, G.~Benato et~al., \emph{Pulse shape discrimination in cupid-mo using principal component analysis}, {\emph{Journal of Instrumentation} {\bfseries 16} (2021) P03032}.

\bibitem{Ahmine:2023xhg}
A.~Ahmine et~al., \emph{{Enhanced light signal for the suppression of pile-up events in Mo-based bolometers for the 0$\nu \beta \beta $ decay search.}}, \href{https://doi.org/10.1140/epjc/s10052-023-11519-6}{\emph{Eur. Phys. J. C} {\bfseries 83} (2023) 373} [\href{https://arxiv.org/abs/2302.13944}{{\ttfamily 2302.13944}}].

\bibitem{golwala2000exclusion}
S.R.~Golwala, \emph{Exclusion limits on the WIMP-nucleon elastic-scattering cross-section from the Cryogenic Dark Matter Search}, University of California, Berkeley (2000).

\end{thebibliography}
